\documentclass[
 aps,
 10pt,
 final,
 notitlepage,
 oneside,
 onecolumn,
 nobibnotes,
 nofootinbib,
 superscriptaddress,
 noshowpacs,
 centertags]
{revtex4}

\def\squareforqed{\hbox{\rlap{$\sqcap$}$\sqcup$}}

\def\sq{\ifmmode\squareforqed\else{\unskip\nobreak\hfil
\penalty50\hskip1em\null\nobreak\hfil\squareforqed
\parfillskip=0pt\finalhyphendemerits=0\endgraf}\fi}

\def\arcmin{\hbox{$^\prime$}}

\def\arcsec{\hbox{$^{\prime\prime}$}}

\def\utw{\smash{\rlap{\lower5pt\hbox{$\sim$}}}}

\def\udtw{\smash{\rlap{\lower6pt\hbox{$\approx$}}}}

\def\diameter{{\ifmmode\mathchoice
{\ooalign{\hfil\hbox{$\displaystyle/$}\hfil\crcr
{\hbox{$\displaystyle\mathchar"20D$}}}}
{\ooalign{\hfil\hbox{$\textstyle/$}\hfil\crcr
{\hbox{$\textstyle\mathchar"20D$}}}}
{\ooalign{\hfil\hbox{$\scriptstyle/$}\hfil\crcr
{\hbox{$\scriptstyle\mathchar"20D$}}}}
{\ooalign{\hfil\hbox{$\scriptscriptstyle/$}\hfil\crcr
{\hbox{$\scriptscriptstyle\mathchar"20D$}}}}
\else{\ooalign{\hfil/\hfil\crcr\mathhexbox20D}}%
\fi}}

\begin{document}

\title{Multi-frequency quasi-simultaneous observations of six very low synchrotron peaked blazars}
\author{\firstname{T.V.}~\surname{Mufakharov}}
\affiliation{Special Astrophysical Observatory of RAS, Nizhnij Arkhyz, 369167 Russia}
\email{timur.mufakharov@gmail.com}

\author{\firstname{Yu.V.}~\surname{Sotnikova}}
\affiliation{Special Astrophysical Observatory of RAS, Nizhnij Arkhyz, 369167 Russia}

\author{\firstname{M.G.}~\surname{Mingaliev}}
\affiliation{Special Astrophysical Observatory of RAS, Nizhnij Arkhyz, 369167 Russia}
\affiliation{Kazan Federal University, 18 Kremlyovskaya St., Kazan, 420008, Russia}

\author{\firstname{A.K.}~\surname{Erkenov}}
\affiliation{Special Astrophysical Observatory of RAS, Nizhnij Arkhyz, 369167 Russia}

\begin{abstract}
We have made an estimation of the
synchrotron peak frequency ($\nu_{peak}^{s}$)
for six very low synchrotron peaked (VLSP) blazars.
These objects were selected as VLSP candidates (with the $\nu_{peak}^{s} \leq 10^{13}$ Hz)
from the archival data.
We have build spectral energy distribution 
 using quasi-simultaneous observations at the
Zeiss-1000 and RATAN-600 telescopes
of the Special astrophysical observatory of RAS
and made an estimation of the $\nu_{peak}^{s}$.
We confirmed classification as a VLSP for two sources 
([HB89]\,1308+326 and 3C\,345), 
for four other blazars we have calculated 
$\nu_{peak}^{s}>10^{13}$ Hz.
\end{abstract}

\maketitle
{\small Keywords: {quasars: general---BL Lacertae objects: general---galaxies: nuclei---galaxies: jets---
radio continuum: galaxies}\par}

\textit{Accepted for publication in Astrophysical Bulletin, Volume 70, Issue 3, 2015}

\section{Introduction}
Blazars form up relatively small ($\sim$ 1\%) sub-group of active galactic nuclei (AGNs),
characterised with a relativistic jet, viewed at small angle by observer on Earth~\cite{1995PASP..107..803U}.
There are two components in their broadband spectral energy distribution (SED):
low-frequency one, with the maximum in NIR/optical/X-ray band,
and high-frequency one, with the peak in gamma-rays.
Those two features generally explained in terms of
synchrotron and inverse Compton mechanisms of emission~\cite{1996ApJ...463..444S};
corresponding peak frequencies indicated with $\nu_{peak}^{s}$ and $\nu_{peak}^{IC}$.
Blazars divided into high/low-frequency peaked, according to the $\nu_{peak}^{s}$:
high synchrotron peaked~(HSPs) ones with the $\nu_{peak}^{s}>10^{16.5}$ Hz,
and low synchrotron peaked~(LSPs) blazars with the $\nu_{peak}^{s}<10^{14.5}$ Hz~\cite{1995ApJ...444..567P}.
Ant\'{o}n and Brown~\cite{2005MNRAS.356..225A} also considered very low energy peaked BL Lacs (VLBL),
with the synchrotron component peaking in IR/mm band $\nu_{peak}^{s} \leq 10^{13}$ Hz.

According to the works on synchrotron peak frequency estimation,
only about 10\% of blazars have $\nu_{peak}^{s}<10^{13}$ Hz,
especially they are rare among BL Lacs.
For example, in~\cite{2006A&A...445..441N} only 3\% of their sample could be classified as VLBL.
According to the BLcat~\cite{2014A&A...572A..59M} 
(includes more than 300 BL Lacertae type objects, observed with the RATAN-600 radio telescope),
about 9\% are VLBL among BL Lacs. 
The typical value of the $\nu_{peak}^{s}$ for FSRQs (flat-spectrum radio quasars)
is $10^{13.1~\pm~0.1}$ Hz, for BL Lacs $\nu_{peak}^{s}$ value is one order of magnitude greater~\cite{2012A&A...541A.160G}.

Nowadays many ground-based and space telescopes observing AGNs
during different monitoring programmes,
also in order to build broadband SED.
It is crucial to analyse quasi-simultaneous SED
because of variable nature of blazars.
The shape and positions of the components of SED could change
during active state of the blazars, which will lead to misclassification.
For example, synchrotron peak frequency 
shifted from $10^{12}$ to $10^{14}$ Hz during the flare
in a gamma-ray band for object 4C\,+49.22~\cite{2014MNRAS.445.4316C}. 
This object could be considered as a transition type, evolving from FSRQ to BL Lac~\cite{2012MNRAS.420.2899G,2013MNRAS.431.1914G}.
A less dramatic case of $\nu_{peak}^{s}$ value change was detected in PKS\,1510-089,
that demonstrated shift from $1.5 \times 10^{13}$ to $6.5 \times 10^{13}$ Hz~\cite{2011A&A...529A.145D}.
 
In this paper we will use quasi-simultaneous observations
in optical (B, V, R bands) and radio (4.8--21.7 GHz) domain
to estimate $\nu_{peak}^{s}$ values for six blazars, 
preliminarily defined by us as a candidates to the very low synchrotron peaked sources.
Throughout this paper we will refer to our objects as very low synchrotron peaked blazars --- VLSP. 

\section{The sample of VLSP blazars}
The blazars monitoring list at the RATAN-600 radio telescope~\cite{1972IzPul.188....3K, 1972IzPul.188...13P, 1976RaF....19.1581P} includes different types of objects (FSRQs and BL Lacs),
the BL Lacertae type objects list could be find in the on-line catalogue\footnote{\url{http://www.sao.ru/blcat/}}~\cite{2014A&A...572A..59M}.
We selected six blazars --- VLSP candidates from our list to observe them with the Zeiss-1000 telescope at the same time with the RATAN-600 observations during year 2014.
We made a preliminarily estimation of the $\nu_{peak}^{s}$ using the ASDC SED Builder Tool\footnote{\url{http://tools.asdc.asi.it}}.
It is useful tool to create multi-waveband SED and made an polynomial approximation
of the different regions of the SED to calculate the $\nu_{peak}^{s}$ value.
In Table~\ref{sample} we present:
source name, right ascension (RA) and declination (DEC) (for the J2000.0 epoch),
redshift value\footnote{data from the NASA/IPAC (NED) \url{http://ned.ipac.caltech.edu/}},
and preliminarily estimated $\nu_{peak}^{s}$ value.

\begin{table*}[H!]
\centering
\setcaptionmargin{0mm} 
\onelinecaptionsfalse
\captionstyle{normal} 
\caption{Sample of the VLSP candidates}
\label{sample}
\medskip
\begin{tabular}{|c|c|c|c|c|}
\hline
Name & RA~(2000.0)& DEC~(2000.0) & $z$ & log~$\nu_{peak}^{s}$ \\
\hline
PKS\,0336-01 & 03$^h$39$^m$30.9$^s$ & -01$^{\circ}$46$^{\arcmin}$36$^{\arcsec}$ & 0.852 & 12.68 \\
PKS\,0446+11 & 04$^h$49$^m$07.7$^s$ & +11$^{\circ}$21$^{\arcmin}$29$^{\arcsec}$ & 1.207 & 12.80 \\
PKS\,0528+134 & 05$^h$30$^m$56.4$^s$ & +13$^{\circ}$31$^{\arcmin}$55$^{\arcsec}$ & 2.06 & 11.96 \\
{[HB89]}\,1308+326 & 13$^h$10$^m$28.6$^s$ & +32$^{\circ}$20$^{\arcmin}$44$^{\arcsec}$ & 0.998 & 12.68 \\
3C\,345 & 16$^h$42$^m$58.8$^s$ & +39$^{\circ}$48$^{\arcmin}$37$^{\arcsec}$ & 0.592 & 12.78 \\
PKS\,2230+11 & 22$^h$32$^m$36.4$^s$ & +11$^{\circ}$43$^{\arcmin}$51$^{\arcsec}$ & 1.037 & 12.86 \\
\hline
\end{tabular}
\end{table*}

\section{Observations and data reduction}
\subsection{Radio data}
We used the flux densities measured at five frequencies (2.3, 4.8, 7.7, 11.2 and 21.7 GHz) 
in 2014 February, March, July and December at the RATAN-600 radio telescope.
Each source was observed from 5 to 12 times during this period\footnote{but for the {[HB89]}\,1308+326 measurements available only at 7.7 GHz}.
The experimental data were processed with the modules of the FADPS
(Flexible Astronomical Data Processing System) standard reduction
package by \cite{1997ASPC..125...46V}. The processing
methods are described in the paper by \cite{2012A&A...544A..25M}.
The following 12 flux density calibrators (standard and
RATAN's traditional ones) were used to
get the coefficients of antenna elevation: 3C48,
3C138, 3C147, 3C161, 3C286, 3C295, 3C309.1, NGC7027,
J0237$-$23, J1154$-$35, J1347$+$12 and J0410$+$76.
Measurements of some calibrators were corrected for angular
size and linear polarization, following the data summarized in \cite{1994A&A...284..331O}
and \cite{1980A&AS...39..379T}, respectively.
The detection limit for the RATAN-600 single sector
is approximately 8~mJy (integration time is about 3 s) under good weather conditions
at the frequency of 4.8 GHz and at an average antenna elevation ($\delta\sim$ $42^{\circ}$).
Standard errors in determining the flux density for these data are
3--5\% at 4.8 и 7.7 GHz,  
4--10\% at 11.2 GHz, and 
10--14\% at 21.7 GHz.
All sources had relatively strong flux levels
 at radio frequencies with a signal-to-noise ratio $S/N \geq 6$.
In Table \ref{table:radio}, we list the radio flux densities at different frequencies used in this work.

\begin{table*}[]
\centering
\setcaptionmargin{0mm} 
\onelinecaptionsfalse
\captionstyle{normal}  
\caption{The flux density values at four frequencies measured with the RATAN-600} 
\label{table:radio}
\medskip
\small
\begin{tabular}{|c|c|c|c|c|c|}
\hline
Name & Date & $F_{21.7 GHz}$, Jy & $F_{11.2 GHz}$, Jy & $F_{7.7 GHz}$, Jy & $F_{4.8 GHz}$, Jy \\
\hline
PKS\,0336-01 & 2014 Dec 7--12 & 1.570 $\pm$ 0.194 & 1.860 $\pm$ 0.180 &  -- & 2.23 $\pm$ 0.096\\
PKS\,0446+11 & 2014 Feb 12 -- Mar 8 & 0.634 $\pm$ 0.078 & 0.782 $\pm$ 0.031 &  -- & 0.694 $\pm$ 0.021\\
PKS\,0528+134 & 2014 Feb 12 -- Mar 8 & 0.811 $\pm$ 0.116 & 1.193 $\pm$ 0.048 &  -- & 1.424 $\pm$ 0.043\\
{[HB89]}\,1308+326 & 2014 May 27--31 & 1.74 $\pm$ 0.07 & 2.14 $\pm$ 0.11 &  2.14 $\pm$ 0.11 & 1.92 $\pm$ 0.23\\
3C\,345 & 2014 July 7--12 & 5.669 $\pm$ 0.538 & 5.704 $\pm$ 0.285 &  -- & 5.846 $\pm$ 0.234\\
PKS\,2230+11 & 2014 Dec 7--12 & 2.571 $\pm$ 0.315 & 3.260 $\pm$ 0.241 &  -- & 4.207 $\pm$ 0.181\\
\hline
\end{tabular}
\normalsize
\end{table*}

\subsection{Optical data}
Photometric optical observations were performed in
2014 February, May, July and November at 1-meter Zeiss-1000 telescope,
equipped with liquid nitrogen-cooled CCD camera EEV 42-40 (2048~$\times$~2048 pixels).
A standard \textit{B}, \textit{V} (Johnson) and \textit{R} (Cousins) filters were used.
Exposure times, depending upon the brightness of the source
and weather conditions, were 30--300 s long\footnote{unfortunately the PKS\,0528+134 was too faint in \textit{B} and \textit{V} filters to estimate its magnitude}.
The average size of the objects was 2--3\arcsec under good astroclimatic conditions.
All the data reduction procedures were conducted using standard methods in the MaxImDL.
Reference stars were chosen with a brightness comparable to the
target star, located in the same field of the CCD matrix. Thus, atmospheric extinction 
was the same for the reference and target star and did not require additional accounting.
Correction for interstellar extinction was made using values from NED.
The standard errors in determining the magnitude are
0.8\% in \textit{B}, 0.7\% in \textit{V}, and 0.4\% in \textit{R} filter.
We used standard formula to convert magnitudes to flux densities: 
\begin{equation*}
m_1-m_2=-2.5~log\frac{f_1}{f_2},
\end{equation*}
there 
$m_1, m_2$ --- are magnitudes of the target and reference stars,
$f_1, f_2$ --- their flux densities, respectively.
If we take the calibration flux density value ($f_0$),
at which magnitude is equal to zero ($m$ = 0),
i.e. write down 
\begin{equation*}
m_2=0,~f_2=f_0,
\end{equation*}
then
\begin{equation*}
m_1=-2.5~log\frac{f_1}{f_0},
\end{equation*}
and flux density will be
\begin{equation*}
f_1=10^{\frac{-m_1}{2.5}}~f_0.
\end{equation*}
The zeropoint flux density values $f_0$ are taken from~\cite{1979PASP...91..589B}
and presented in Table~\ref{table:bvr} for each filter.
The results of the photometry are presented in Table~\ref{table:optic}.

\begin{table}[]
\centering
\setcaptionmargin{0mm} 
\onelinecaptionsfalse
\captionstyle{normal}  
\caption{Parameters of the optical filters and their calibration flux density values ($f_0$ for mag = 0) from~\cite{1979PASP...91..589B}}
\label{table:bvr}
\medskip
\begin{tabular}{|c|c|c|c|}
\hline
 Band & $\lambda$,~nm & $\nu$,~Hz & $f_0$,~Jy  \\
\hline
B  & 440 & $6.81\times10^{14}$ & 4260\\
V  & 550 & $5.45\times10^{14}$  & 3640 \\
R  & 640 & $4.68\times10^{14}$  & 3080 \\
\hline
\end{tabular}
\end{table}

\begin{table*}[]
\centering
\setcaptionmargin{0mm} 
\onelinecaptionsfalse
\captionstyle{normal}  
\caption{Magnitudes at three bands, measured with the Zeiss-1000, and corresponding flux densities of the objects from our sample (this values before correction for interstellar extinction)}
\label{table:optic}
\medskip
\small
\begin{tabular}{|c|c|c|c|c||c|c|c|}
\hline
Name & Date & B, mag & V, mag & R, mag & B, mJy & V, mJy & R, mJy  \\
\hline
PKS\,0336-01 & 2014 Nov 19 & 17.65 $\pm$ 0.12 & 17.54 $\pm$ 0.09 & 17.21 $\pm$ 0.05 & 0.36 $\pm$ 0.0025 & 0.364 $\pm$ 0.0018 & 0.384 $\pm$ 0.0012\\
PKS\,0446+11 & 2014 Feb 2 & 18.14 $\pm$ 0.15 & 18.16 $\pm$ 0.13 & 18.61 $\pm$ 0.10 & 0.229 $\pm$ 0.0019 & 0.206 $\pm$ 0.0015 & 0.106 $\pm$ 0.0006\\
PKS\,0528+134 & 2014 Feb 2 & -- & -- & 19.27 $\pm$ 0.08 & -- & -- & 0.0576 $\pm$ 0.0002\\
{[HB89]}\,1308+326 & 2014 May 28 & 18.64 $\pm$ 0.09 & 18.33 $\pm$ 0.12 & 17.70 $\pm$ 0.05 & 0.145 $\pm$ 0.0007 & 0.176 $\pm$ 0.001 & 0.245 $\pm$ 0.0007\\
3C\,345 & 2014 July 24 & 18.37 $\pm$ 0.26 & 18.18 $\pm$ 0.20 & 17.76 $\pm$ 0.07 & 0.185 $\pm$ 0.0026 & 0.202 $\pm$ 0.0023 & 0.231 $\pm$ 0.0009\\
PKS\,2230+11 & 2014 Nov 19 & 17.39 $\pm$ 0.10 & 16.78 $\pm$ 0.07 & 16.50 $\pm$ 0.06 & 0.457 $\pm$ 0.0025 & 0.734 $\pm$ 0.0031 & 0.739 $\pm$ 0.0026\\
\hline
\end{tabular}
\normalsize
\end{table*}

\section{Results}
\begin{figure*}[]
\centering
\begin{tabular}{cc}
\begin{minipage}{0.45\linewidth}
\center{\includegraphics[width=\textwidth]{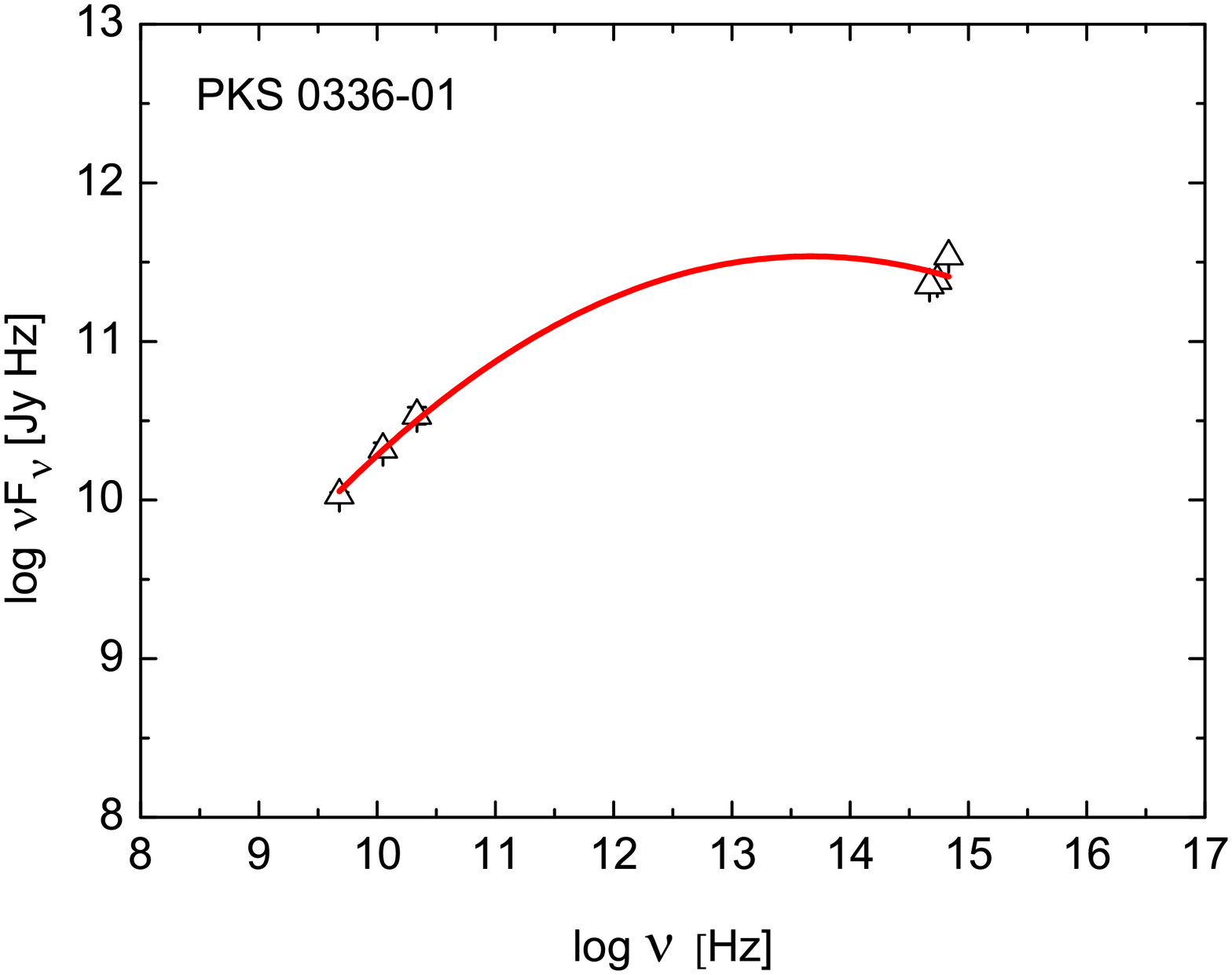}} \\ 
\end{minipage}
\hfill
\begin{minipage}{0.45\linewidth}
\center{\includegraphics[width=\textwidth]{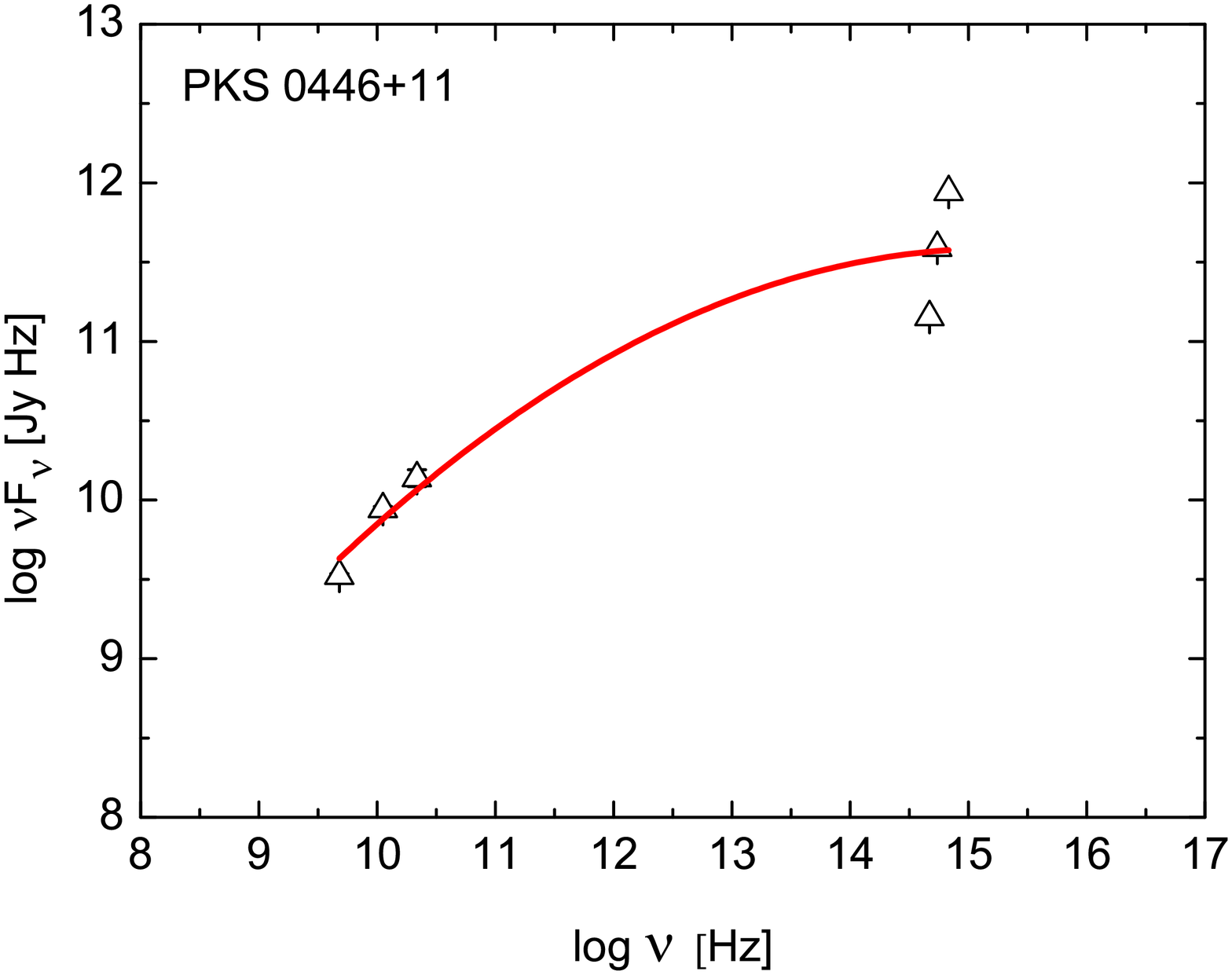}} \\ 
\end{minipage}
\\
\\
\begin{minipage}{0.45\linewidth}
\center{\includegraphics[width=\textwidth]{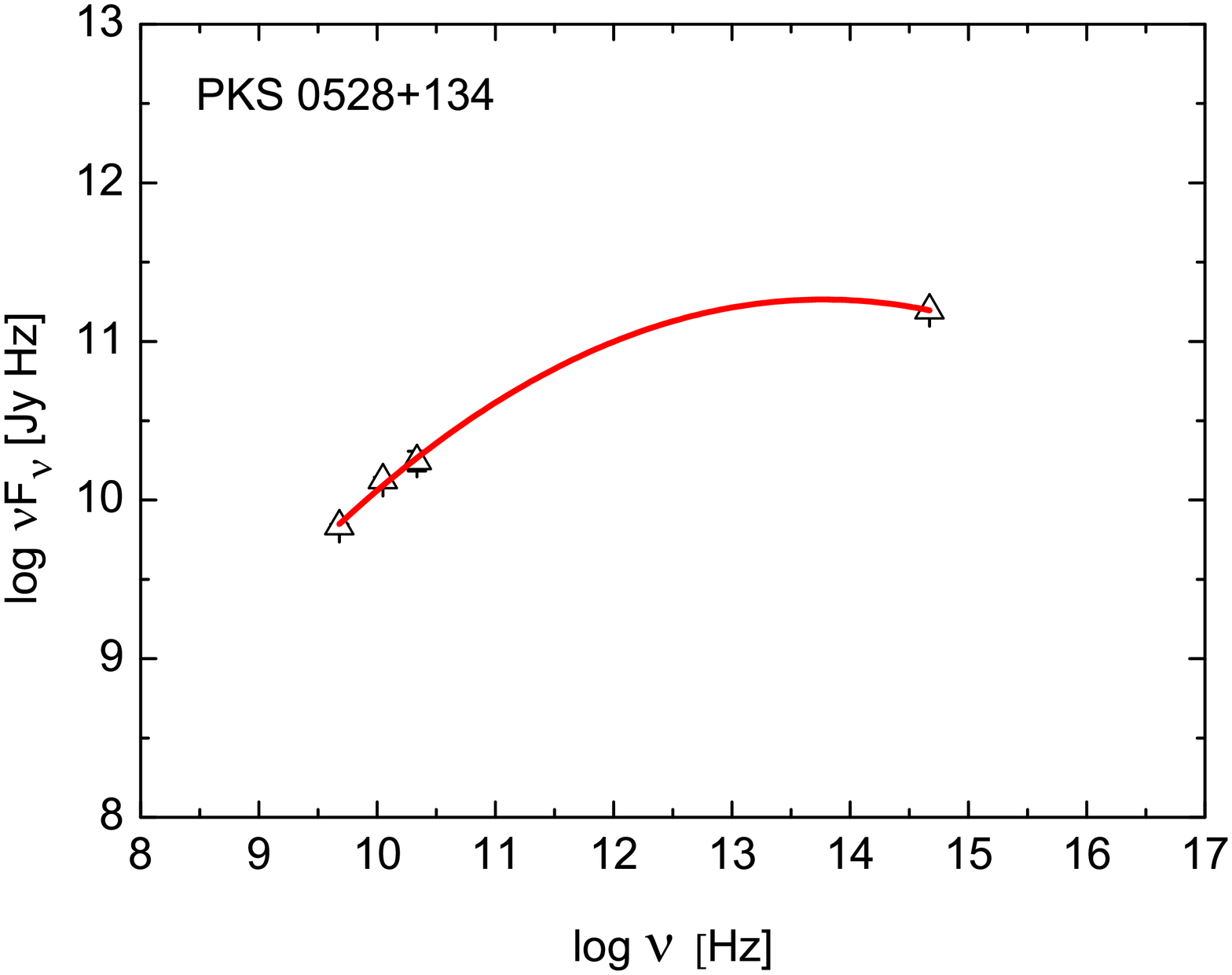}} \\ 
\end{minipage}
\hfill
\begin{minipage}{0.45\linewidth}
\center{\includegraphics[width=\textwidth]{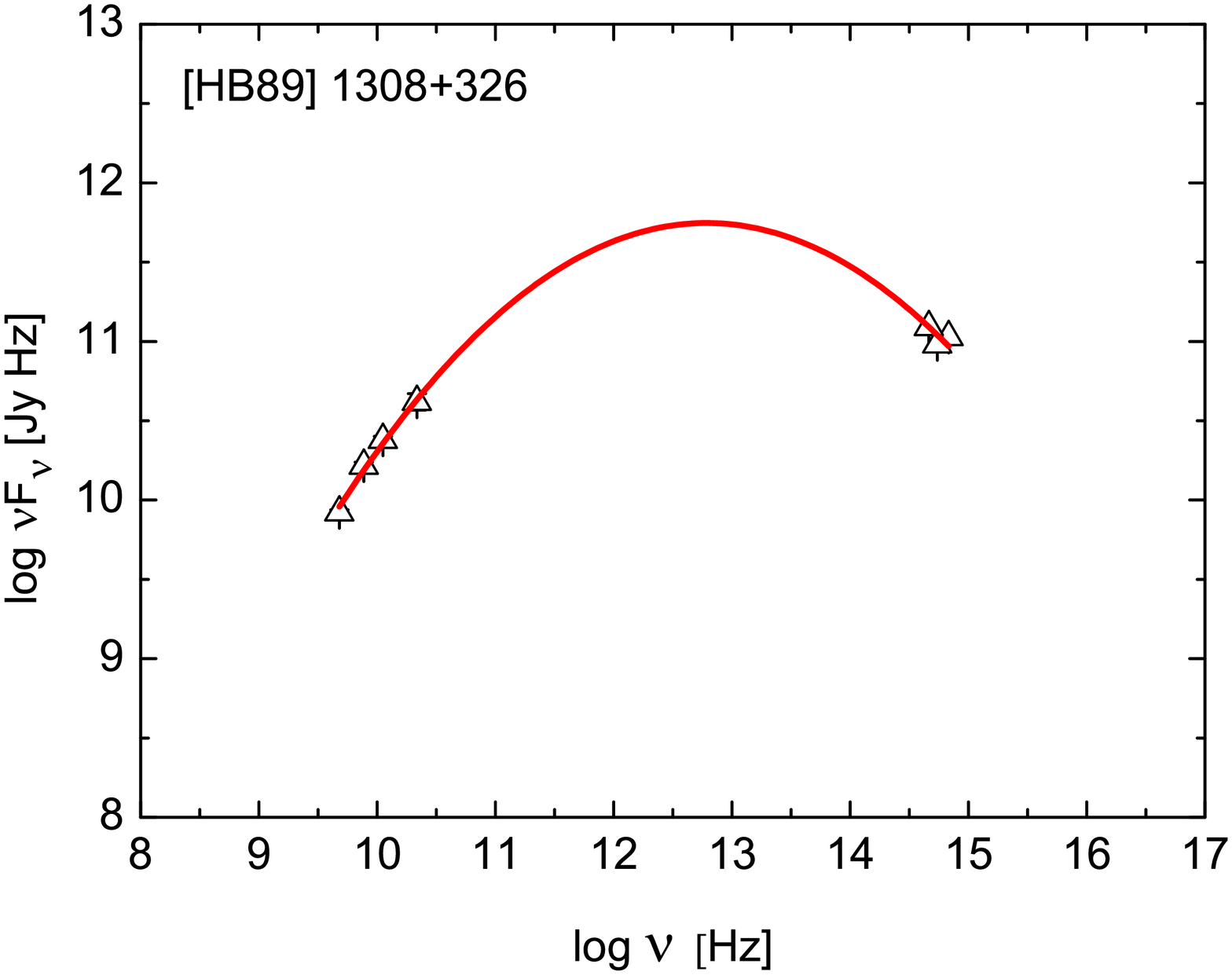}} \\ 
\end{minipage}
\\
\\
\begin{minipage}{0.45\linewidth}
\center{\includegraphics[width=\textwidth]{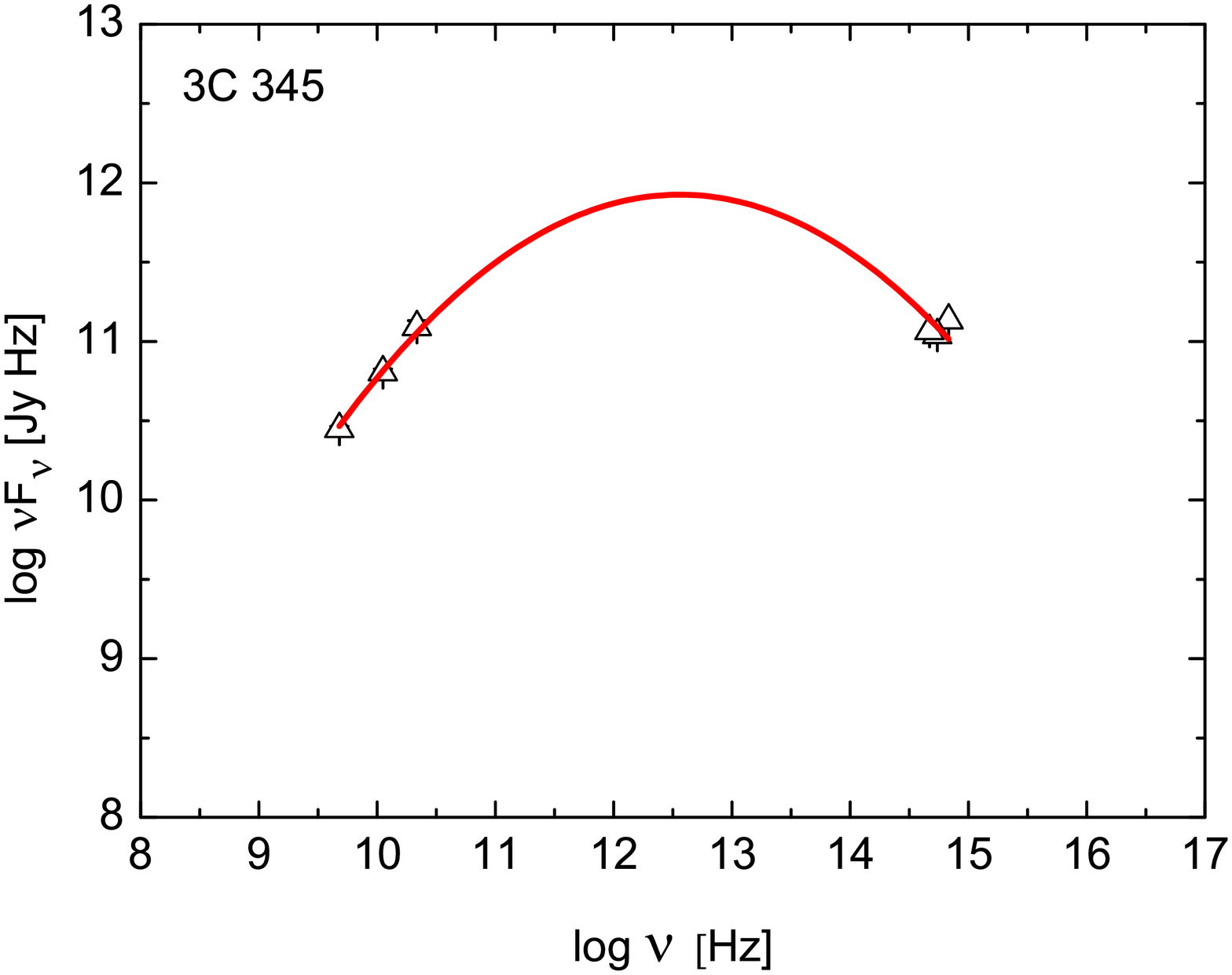}} \\ 
\end{minipage}
\hfill
\begin{minipage}{0.45\linewidth}
\center{\includegraphics[width=\textwidth]{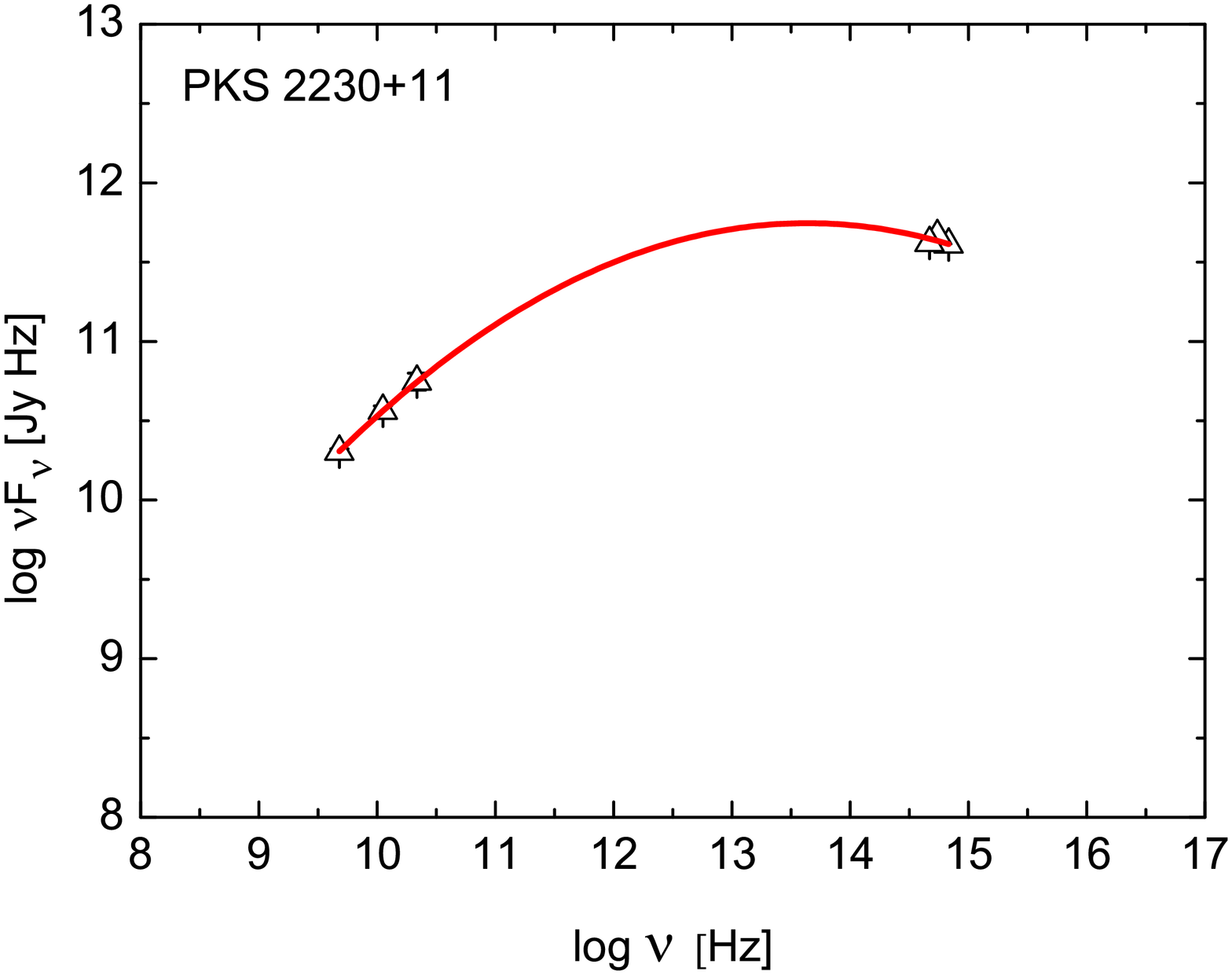}} \\ 
\end{minipage}
\end{tabular}
\caption{Spectral energy distributions build using our RATAN-600 and Zeiss-1000 measurements for objects from our sample.}
\label{fig1}
\end{figure*}

We plotted spectral energy distributions of our VLSP candidates
 using our measurements in radio and optical domain (flux densities presented in Table \ref{table:radio} and Table \ref{table:optic})
in Fig \ref{fig1}.
Most of the observed radio to optical (and in some cases X-ray) emission from blazars is
due to synchrotron radiation of the relativistic electrons moving in a magnetic field \cite{1981Natur.293..714B,1982ApJ...253...38U,1988AJ.....95..307I,1998ASPC..144...25M}.
Its spectrum described by a power low \cite{1979rpa..book.....R} 
and second or third degree polynomial is usually used to describe form of the synchrotron component of the spectrum (for example to find its maximum) \cite{1995MNRAS.277..297C,1998ApJ...504..693K,2006A&A...445..441N,2010ApJ...716...30A,2011A&A...536A..15P}.

The synchrotron component of the SED ($10^8 - 10^{15}$ Hz) in our case was fitted with a parabolic function:
\begin{equation*}
log~(\nu F) = A + B (log~\nu) + C (log~\nu)^2
\end{equation*}
We used the OriginLab (software for data analysis and graphing) to estimate synchrotron peak frequency.
The results of $\nu_{peak}^{s}$ estimation based on our almost concurrent and archival non-concurrent data
for our sample are presented in Table \ref{table:result}.
Comments on individual sources provided below. 

\begin{table*}[]
\centering
\setcaptionmargin{0mm} 
\onelinecaptionsfalse
\captionstyle{normal} 
\caption{Measurements of the $\nu_{peak}^{s}$ parameter for the investigated objects from our sample (at column two, designated SAO), 
estimated values of the $\nu_{peak}^{s}$ using SED Builder Tool also provided (at column three, designated archive). Objects, confirmed as VLSP, marked with + (at column four)}
\label{table:result}
\medskip
\begin{tabular}{|c|c|c|c|}
\hline
Name & log~$\nu_{peak}^{s}$, SAO & log~$\nu_{peak}^{s}$, archive & VLSP \\
\hline
PKS\,0336-01 & 13.67 & 12.68 & - \\
PKS\,0446+11 & 15.30 & 12.80 & - \\
PKS\,0528+134 & 13.77 & 11.96 & - \\
{[HB89]}\,1308+326 & 12.79 & 12.68 & +\\
3C\,345 & 12.55 & 12.56 & +\\
PKS\,2230+11 & 13.64 & 12.86 & - \\
\hline
\end{tabular}
\end{table*}

\textbf{PKS\,0336-01}.
This source has been classified as a flat-spectrum radio quasar (FSRQ)~\cite{2007ApJS..171...61H},
with the dominating radio core.
The optical light curve presented at the St.~Petersburg State University (SPbSU)
 virtual observatory\footnote{\url{http://lacerta.astro.spbu.ru/}}
shows decreasing of the \textit{R} band magnitude at the end of the 2014: from $\sim$15.7 to $\sim$17.2.
In October 2014 a GeV gamma-ray flare of this object was detected~\cite{2014ATel.6568....1C},
but at the same time there was no activity in optical band~\cite{2014ATel.6577....1N}.
Interestingly, after couple of weeks NIR flare of this blazar was detected~\cite{2014ATel.6662....1C}.
According to the Owens Valley Radio Observatory (OVRO) monitoring\footnote{\url{http://www.astro.caltech.edu/ovroblazars/data/data.php}},
source was stable at the radio band at the end of 2014 (flux density at 15 GHz $\sim$2.4 Jy).

Our observations were performed in November 19 and in December 7--12 in optical and radio band, respectively.
We determined log~$\nu_{peak}^{s}$ = 13.67
(four points, the reduced chi-square $\chi^2$ = 0.0092), which is somewhat higher than archival
value log~$\nu_{peak}^{s}$ = 12.68.

\textbf{PKS\,0446+11}.
This is BL Lacertae type object~\cite{2006A&A...455..773V}, with almost featureless optical spectra~\cite{2003AJ....125..572H},
have a flat radio spectrum~\cite{2007ApJS..171...61H}.
The source was in a bright state at radio band in early 2014, according to OVRO monitoring programme.
Note that the radio light curve of this source characterized with a slow rise (and decline) during activity,
maximum of a brightness is reached usually in a scale of about one year.
The flux density variation is less than 10\% in a scale of 1--2 weeks.

Our observations were performed in February 2 in optical band and during February 12 -- March 8
in radio band.
From the SED of this object (see Fig~\ref{fig1})
it could be seen that the positions of the points in optical band
influenced the parabolic fit, so that 
maximum became higher than it could be expected.
We determined log~$\nu_{peak}^{s}$ = 15.30 (six points, $\chi^2$ = 0.108),
while estimated value from archival data is log~$\nu_{peak}^{s}$ = 12.80.

\textbf{PKS\,0528+134}.
This blazar has a radio jet~\cite{2002A&A...381..757L} and 
classified as a FSRQ~\cite{2007ApJS..171...61H}. 

We observed this source in February 2 in optical band and during February 12 -- March 8 in radio band.
This source was generally very faint in optical band, and we obtained only three images in \textit{R} filter
(3 expositions, 300 s each), we failed to accumulate sufficient amount of photons in \textit{B} and \textit{V} filters to calculate its magnitude.
Later in that year, in 2014 November, we tried again but it was still in a faint state (less than 19 mag in \textit{R} filter).
We estimated log~$\nu_{peak}^{s}$ = 13.77 
(four points, $\chi^2$ = 0.0017);
according to the archival data log~$\nu_{peak}^{s}$ = 11.96.

\textbf{[HB89]\,1308+326}.
It is BL Lacertae type object, with weak emission lines, 
host galaxy is not detectable~\cite{1993A&AS...98..393S,2005A&A...440..831K}. 
A strong flare was detected in gamma-ray band in the 2014 April~\cite{2014ATel.6068....1S},
but following optical observations did not reveal activity~\cite{2014ATel.6072....1N}.
At the first half of 2014 this blazar was in active state in radio band
(flux density at 15 GHz $\sim$2 Jy, according to the OVRO monitoring programme).

We observed this source in May 28 and five times in May 27--31 in optical and radio band, respectively.
We estimated log~$\nu_{peak}^{s}$ = 12.79 (seven points, $\chi^2$ = 0.0025),
which is in a good agreement with the value obtained using archival data --- 
 log~$\nu_{peak}^{s}$ = 12.68.

\textbf{3C\,345}.
This source has been classified as a FSRQ \cite{2007ApJS..171...61H},
with the dominating radio core \cite{1983MNRAS.204..151L}.
 It was in a quiescent state in optical band in 2014, slightly brighter than 18 mag in \textit{R} filter
 (according to the SPbSU virtual observatory);
 slight fall in brightness in radio domain, which began in 2013, has continued --- from $\sim$6 to $\sim$5 Jy (OVRO data).
 
 Our observations were performed in July 24 in optical band and two weeks earlier, in July 7--12 in radio band.
We estimated log~$\nu_{peak}^{s}$ = 12.56 (four points, $\chi^2$ = 0.0079).
The log~$\nu_{peak}^{s}$ = 12.78, according to the archival data.
It is the second source in our sample for which our estimated value of the synchrotron peak 
is very close to that determined from archival non-concurrent data. 

\textbf{PKS\,2230+11}.
This source is known as a typical blazar \cite{2007A&A...464..175B}
and as a quasar with the high optical polarization (>3\%)~\cite{2006A&A...455..773V};
a radio jet is also observed in the object \cite{2002A&A...381..757L}.
The significant SED variation was detected during multi-waveband monitoring of its flare in 2005 \cite{2007A&A...464..175B}.
One can approximately estimate the $\nu_{peak}^{s}$ value decrease of about one order,
 with the spread of activity to the low-frequency region of the spectrum,
  from the SED presented in that work \cite{2007A&A...464..175B}.
There was a brightening of about one order of magnitude in \textit{R} band at the end of the 2014, 
according to the light curve presented in the SPbSU virtual observatory web page
 (in comparison with its stable state in the middle of 2014).
It is possible that this activity is linked with the flare in a gamma-ray band, 
registered in late October of the same year \cite{2014ATel.6631....1C}.

We observed this source in November 19 in optical band and three weeks later, during December 7--12, in radio band.
We determined log~$\nu_{peak}^{s}$ = 13.64 (four points, $\chi^2$ = 0.0005);
log~$\nu_{peak}^{s}$ = 12.86, archival data.

\section{Conclusions}

We have presented new observational results at radio and optical bands
for six blazars, which were selected to be VLSP candidates ($\nu_{peak}^{s}<10^{13}$ Hz)
 according to the non-concurrent archival data. 
We estimated synchrotron peak frequency $\nu_{peak}^{s}$ 
using quasi-simultaneous observations at the Zeiss-1000 and the RATAN-600 telescopes,
the difference in observation epochs was not more than 2--3 weeks.

We confirmed classification as VLSP for two sources:  [HB89]\,1308+326 и 3C\,345.
Other four objects considered as candidates to VLSP (PKS\,0336-01, PKS\,0446+11, PKS\,0528+134 and PKS\,2230+11)
 have shown $\nu_{peak}^{s}>10^{13}$ Hz.
We found synchrotron peak frequency values very close to the estimated from archival data
for [HB89]\,1308+326 and 3C\,345.

\begin{acknowledgments}
The authors want to acknowledge O.~Spiridonova, M.~Gabdeev and A.~Moskvitin
for their help with the Zeiss-1000 observations.
The RATAN-600 observations were carried out with the
financial support of the Ministry of Education and Science of the Russian
Federation.
The authors acknowledges support through the Russian Government Program of Competitive Growth of the Kazan Federal University.
We used on-line version of the Roma-BZCAT catalogue and the SED Builder Tool at the ASI Science Data Center (ASDC) website,
therefore we are grateful to the ASDC staff.
This research has made use of the NASA/IPAC Extragalactic Database (NED) which is operated by the Jet Propulsion Laboratory, California Institute of Technology, under contract with the National Aeronautics and Space Administration.
\end{acknowledgments}

\bibliographystyle{AstroBull} 
\bibliography{biblioteka} 

\end{document}